\documentclass[a4,12pt]{article}
\usepackage{a4wide}
\usepackage{amssymb}
\usepackage{amsmath}
\usepackage{graphicx}
\usepackage{xspace}
\usepackage{epsfig}
\newcommand{\ee}{e^+e^-}
\newcommand{\qq}{q \bar q}
\newcommand{\as}{\alpha_s}
\newcommand{\CA}{C_A}
\newcommand{\TR}{T_R}
\newcommand{\order}[1]{\mathcal{O}\left(#1\right)}
\newcommand{\ie}{i.e.\ }
\newcommand{\eg}{e.g.\ }
\newcommand{\lft}{\mathrm{left}}
\newcommand{\rght}{\mathrm{right}}

\newcommand{\TeV}{\,\mathrm{TeV}}

\begin{document}

%%%%%%%%%%%%%%%%%%%%%%%%%%%%%%%%%%%%%%%%%%%%%%%%%%%%%%%%%%%%%%%%%%%%%%%
\titlepage
\begin{flushright}
hep-ph/0601139 \\
Bicocca-FT-05-28\\
Cavendish-HEP-05/25\\
CERN-PH-TH-06/02\\
DAMTP-2005-134\\
FERMILAB-PUB-06-003-T\\
LPTHE-05-34\\
January 2006\\
%Draft $Revision: 1.60 $
\end{flushright}

\vspace*{0.3in}
\begin{center}
  {\Large \textbf{\textsf{ Infrared safe definition of jet flavour
      }}}\\
  \vspace*{0.4in} Andrea~Banfi$^{(a,b,c)}$, Gavin~P.~Salam$^{(d)}$
  and Giulia~Zanderighi$^{(e,f)}$ \\
  {\small \vspace*{0.5cm}
    $^{(a)}$ {\it DAMTP, Centre for Mathematical Sciences, Cambridge CB3 0WA, UK}. \\
    \vskip 2mm
    $^{(b)}$ {\it Cavendish Laboratory, University of Cambridge, Cambridge CB3 0HE, UK}. \\
    \vskip 2mm
    $^{(c)}$ {\it University of Milano--Bicocca and INFN Sezione di Milano, 20126 Milan, Italy}. \\
    \vskip 2mm $^{(d)}$ {\it LPTHE: Universit\'e Pierre
      et Marie Curie --- Paris~6; \\
      Universit\'e Denis Diderot --- Paris~7;
      CNRS;
      75252 Paris 75005, France.}\\
    \vskip 2mm $^{(e)}$ {\it Theory Group, Fermilab, P.O. Box 500,
      Batavia, IL, US.
    }\\
    \vskip 2mm} $^{(f)}$ {\it Theory Division, Physics Department,
    CERN, 1211 Geneva 23, Switzerland.
  }\\
  \vskip 2mm
\end{center}

\vspace{0.5cm}
\begin{abstract}
  It is common, in both theoretical and experimental studies, to
  separately discuss quark and gluon jets.  However, even at parton
  level, widely-used jet algorithms fail to provide an infrared safe
  way of making this distinction. We examine the origin of the
  problem, and propose a solution in terms of a new `flavour-$k_t$'
  algorithm. As well as being of conceptual interest this can be a
  powerful tool when combining fixed-order calculations with multi-jet
  resummations and parton showers. It also has applications
  to studies of heavy-quark jets. 
\end{abstract}
\vspace{0.5cm}
%%%%%%%%%%%%%%%%%%%%%%%%%%%%%%%%%%%%%%%%%%%%%%%%%%%%%%%%%%%%%%%%%%%%%%%

\newpage

%======================================================================
\section{Introduction}
\label{sec:introduction}

A search through the SPIRES database reveals over 350
articles whose titles contain the expressions `quark jet(s)'
or `gluon jet(s)' \cite{Spires}. The idea of quark and gluon jets appears so
intuitive that it hardly seems necessary to examine the question of
what it means. Yet, when going beyond leading order perturbative QCD,
the concept of quark and gluon jets is only meaningful once a
procedure has been defined to classify an ensemble of partons into a
set of jets, each with a well-defined flavour --- a flavour that is
insensitive to the addition of extra soft or collinear branchings. To
our knowledge the question of how to do this in general has not been
addressed in the literature.

As well as being of intrinsic interest, the question of how to
define the flavour of a partonic jet is becoming of increasing
practical importance as the study of QCD is extended to multi-jet
ensembles (by jets we mean both incoming and outgoing ones): in
studies of $\ee \to $~jets one knows that the basic $2$-jet Born
configuration consists of quark jets; but for jet production at hadron
colliders, the Born configuration involves $2$ incoming and $2$ outgoing
jets and many flavour channels are possible, $qq\to qq$, $q\bar q \to
gg$, $gg \to gg$, etc.
The ability to assign flavours to the jets is especially useful when
combining fixed-order predictions with all-order calculations (be it
for parton showers as in \cite{CKKW} or for analytical resummation
\cite{KOS,BCMN,caesar}).  This is because all-order calculations are
carried out for a fixed Born configuration, with a single flavour
channel at a time, while fixed-order calculations implicitly sum over
all flavour channels and can at best be split up a posteriori to match onto
the individual flavour channels of the all-order calculation.

\begin{figure}[htbp]
  \centering
  \includegraphics[width=0.95\textwidth]{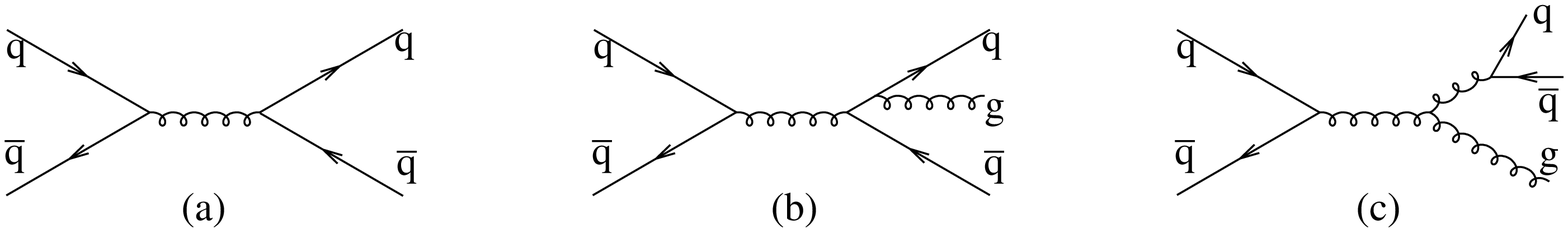}
  \caption{(a) Specific $q\bar q \to q\bar q$ flavour channel for a
    $2\to2$ parton scattering process; (b) higher-order diagram that
    can be seen as a correction to (a); (c) higher-order diagram that
    can be seen as a correction to the process $q\bar q \to gg$, but
    with the same final-state partons as (b).}
  \label{fig:2to3}
\end{figure}

As a concrete example, consider the calculation of higher-order
corrections to the process $q\bar q \to q\bar q$,
fig.~\ref{fig:2to3}a.  An all-order calculation treats the addition of
any number of soft/collinear gluons and extra $q\bar q$ pairs
implicitly, leaving the underlying $2\to 2$ flavours unchanged. When
trying to supplement this with results of a fixed order calculation
one encounters the problem that higher-order contributions cannot be
uniquely assigned to any given $2\to 2$ flavour channel --- the
$\order{\as}$ corrections to $q\bar q \to q\bar q$ include \eg a $q \bar
q\to q \bar q \to q\bar q g$ piece, but a fixed order calculation
gives only the squared sum of all $q\bar q \to q\bar qg$ diagrams,
among them $q\bar q\to q \bar q \to q\bar q g$ and $q\bar q \to g g
\to q\bar qg$, illustrated in fig.~\ref{fig:2to3}b and \ref{fig:2to3}c
respectively.  There can exist no unambiguous procedure for separating
the $q \bar q\to q\bar q g$ contribution into its different underlying
channels, both because the different channels are not individually
gauge invariant and because they interfere when squaring the
amplitude.

One therefore needs a prescription to assign $q \bar q\to q\bar q g$
either to the $q \bar q\to q\bar q$ or the $q \bar q\to gg$ underlying
Born $2\to2$ process (or else to declare it irreducibly $2\to 3$
like), it only being in the $q \bar q\to q\bar q$ case that one needs
to put it together with the $q\bar q \to q\bar q$ all-order
calculation.  This reclassification of a $2\to3$ event as a $2\to2$
event is similar conceptually to what is done in a normal jet
algorithm, except that not only should the momenta of the resulting
$2\to2$ configuration be infrared and collinear safe, but so should
the flavours. Accordingly we call it a jet-flavour algorithm.

An obvious approach to defining jet flavours at the perturbative level
would be to start with an existing jet algorithm, such as the
$k_t$-clustering \cite{ktee,kthh,incl-kthh} or cone \cite{cone}
algorithm, that defines jets such that each particle belongs to at
most one jet. One can then determine the net flavour content of each
of the jets, as the total number of quarks minus antiquarks for each
quark flavour. Jets with no net flavour are identified as gluon jets,
those with (minus) one unit of net flavour are (anti) quark jets,
while those with more than one unit of flavour (or both a flavour and
a different antiflavour) cannot be identified with a single QCD
parton.

\begin{figure}[htbp]
  \centering \includegraphics{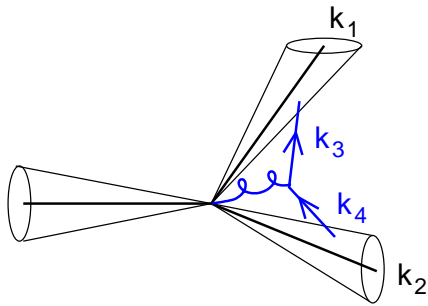}
  \caption{A large-angle soft gluon splitting to a large-angle soft $q\bar q$
    pair ($k_3$, $k_4$) with the $q$ and $\bar q$ then clustered into
    different jets ($k_1$, $k_2$).}
  \label{fig:badjets}
\end{figure}

Applied to the $k_t$ or cone algorithms, this procedure yields a jet
flavour that is infrared (IR) safe at (relative) order $\as$
discussed in our example above. However at (relative) order $\as^2$ a
large-angle soft gluon can split into a widely separated soft $q\bar
q$ pair and the $q$ and $\bar q$ may end up being clustered into
different jets, `polluting' the flavour of those jets, see
fig.~\ref{fig:badjets}. Because this happens for arbitrarily soft
gluons branching to quarks, the resulting jet flavours are infrared
unsafe from order $\as^2$ onwards. We are not aware of this problem
having been discussed previously in the literature, though there do
exist statements that are suggestive of IR safety issues when
discussing flavour \cite{NagySoper}.

In section~\ref{sec:k_t-flav-algor} we shall discuss IR flavour
unsafety with respect
to the $k_t$ (or `Durham') algorithm in $\ee$~\cite{ktee}. There we
shall recall that the $k_t$ closeness measure is specifically related
to the divergences of QCD matrix elements when producing soft and
collinear gluons. However there are no divergences for the production
of soft quarks and, as we shall see, it is the use for quarks of a
distance measure designed for gluons that leads to the infrared
unsafety of jet flavour in the $k_t$ algorithm. By taking into account
the absence of a soft-quark divergence when designing the
jet-clustering distance measure, one can eliminate the infrared
divergence of the jet flavour.

The essence of the modification to the $k_t$ distance is that instead
of the $\min(E_i^2,E_j^2)$ factor that appears usually, one needs to
use $\max(E_i^2,E_j^2)$ when the softer of $i,j$ is a quark. In
section~\ref{sec:flav-algor-hadr} we will examine how this can be
extended to processes with incoming hadrons. There the added
difficulty is the need for a particle-beam distance measure.
Traditionally this involves only one dimensionful scale, related to
the squared transverse-momentum $k_{ti}^2$ of the particle.  There is
a sense in which this can be understood as $\min(k_{ti}^2, k_{tB}^2)$,
where $k_{tB}^2$ is some transverse scale associated with the beam
that is larger than all $k_{ti}^2$ and so could up to now be ignored.
In order to obtain a sensible jet-flavour algorithm we shall however
need to consider also $\max(k_{ti}^2, k_{tB}^2)$ and therefore in
section~\ref{sec:flav-algor-hadr} we shall investigate how to
construct sensible `beam scales'. 

As well as explaining how to build jet algorithms that provide an
infrared safe jet flavour, we shall also examine how they fare in
practice. In $\ee$ it will be possible to carry out tests both with an
NLO code (which explicitly reveals the IR unsafety of flavour in
traditional jet algorithms) and with parton-shower Monte Carlo codes.
For hadron-hadron collisions only parton-shower Monte Carlo tests will
be possible because none of the currently available NLO codes provides
access to the final-state parton flavour information.

%======================================================================
\section{The $\boldsymbol{k_t}$-flavour algorithm for $\boldsymbol\ee$}
\label{sec:k_t-flav-algor}

The aim of clustering algorithms is to recombine particles into jets
in a manner that approximates the inverse of the nearly probabilistic
picture of ordered QCD branching. Since, however, the branching itself
is a quantum mechanical process, there is no unique way of inverting
it for a given final ensemble of particles. What can at most be done
is to design it to work correctly in limits in which the QCD branching
behaves classically, \eg when a given particle is emitted as if
from a single identifiable parent. The design of good jet algorithms
is therefore more a craft than a deductive science.  Nevertheless certain
general principles will help us identify how to extend existing jet
algorithms to deal properly with flavour.

Let us start by considering the most widespread clustering algorithm,
the standard $\ee$ Durham (or $k_t$) algorithm~\cite{ktee}:
\begin{enumerate}
\item Introduce a distance measure $y_{ij}^{(D)}$ between every pair
  of partons $i$, $j$:
  \begin{equation}
    \label{eq:yij}
    y_{ij}^{(D)} = \frac{2\min(E_i^2,E_j^2)}{Q^2} (1 - \cos \theta_{ij})\,,
  \end{equation}
  where $E_i$ is the energy of particle $i$, $\theta_{ij}$ is the
  angle between particles $i$ and $j$ and $Q$ is the centre of mass
  energy. 
\item Find the specific $i$ and $j$ that correspond to the smallest
  $y_{ij}^{(D)}$ and recombine them according to some recombination
  scheme (we shall here use the $E$ scheme, which sums the
  four-momenta).
\item Repeat the procedure until all $y_{ij}^{(D)} > y_{cut}$ (or,
  alternatively, until one reaches a predetermined number of jets).
\end{enumerate}
The defining characteristic of such clustering algorithms is the
distance measure, because it determines the order in which emissions
are recombined.\footnote{There exist also jet-algorithms in which the
  measure that determines the order of recombination differs from that
  defining the stopping point for recombination, \eg the Cambridge and
  Aachen algorithms~\cite{CamAachen}.} %
It is closely related to the divergences of the QCD matrix elements
--- for a gluon $j$ that is soft and collinear to a gluon $i$ the
product of phase-space and matrix element for a parent gluon to branch to
$i$ and $j$ is
\begin{equation}
  \label{eq:soft-col-mat}
  [dk_j] |M^2_{g\to g_i g_j}(k_j)| \simeq \frac{\as\CA}{\pi} \frac{dE_j}{E_j}
  \frac{d\theta_{ij}^2}{\theta_{ij}^2}\,, \qquad (E_j \ll E_i\,, \;\,
  \theta_{ij} \ll 1)\,.
\end{equation}
Thus with the distance measure eq.~(\ref{eq:yij}), two particles are
deemed to be close when either of the parameters in which the matrix
element has a divergence, $E_j = \min(E_i,E_j)$ or $\theta_{ij}$, is
small.  This is a key characteristic of a good distance measure
because where there is a strong divergence there will be many
splittings that are independent of the `hard' properties of the event
--- such splittings should be undone (recombined) at the early stages
of the clustering to leave at the end only well-separated hard
pseudo-jets.

A second key characteristic of a distance measure can be understood by
examining the Jade algorithm~\cite{jade}, which is identical to (and
predates) the $k_t$ algorithm, except that its distance measure is
\begin{equation}
  \label{eq:yij-Jade}
  y_{ij}^{(J)} = \frac{2E_i E_j}{Q^2} (1 - \cos \theta_{ij})\,.
\end{equation}
Again, for $E_j \ll E_i$, $\theta_{ij} \ll 1$, the distance
$y_{ij}^{(J)}$ becomes smaller when either $E_j$ or $\theta_{ij}$ is
reduced, \ie whenever the matrix-element divergence is made stronger.
However it also becomes smaller when $E_i$ is reduced, even though a
modification of $E_i$ has no effect on the divergence of the matrix
element in eq.~(\ref{eq:soft-col-mat}). The undesirable consequence of
this is that the Jade algorithm strongly `prefers' to recombine pairs
of soft particles at large relative angle, instead of combining the
individual soft particles with any collinear but harder neighbours,
and so `pulls' particles out of their natural jet.

From this brief discussion, one can see that the distance measure
should satisfy two main characteristics: (a) two particles should be
considered close when there is a corresponding divergence in their
matrix elements;\footnote{This discussion is somewhat of an
  oversimplification --- for example the Angular-ordered Durham
  algorithm \cite{CamAachen} retains only the angular part of the
  closeness measure and nonetheless behaves sensibly.}  and (b) the
measure should not inadvertently introduce `spurious' extra closeness
for a variation of the momenta that does not lead to any extra
divergence (see however discussion below
eq.~\eqref{eq:yij-flavour-alpha}).

For generic hadron-level jet studies the Durham
measure eq.~(\ref{eq:yij}) is a good choice because the majority of
emissions are gluons --- the correct matrix element to consider in the
design of the measure is that for soft gluon emission (be it
from a quark or a gluon) and it always has both a soft (energy) and
collinear (angular) divergence. For flavour algorithms one
should remember that the matrix elements for $g \to q\bar q$ or $q \to
qg$ (with a soft quark) have no soft divergence, but just the
collinear divergence,
\begin{equation}
  \label{eq:soft-col-mat-g2qq}
  [dk_j] |M^2_{g\to q_i\bar q_j}(k_j)| \simeq \frac{\as\TR}{2\pi}
  \frac{dE_j}{E_i} 
  \frac{d\theta_{ij}^2}{\theta_{ij}^2}\,,\qquad\quad (E_j \ll E_i\,, \;\,
  \theta_{ij} \ll 1)\,,
\end{equation}
(note the index $i$ in the energy
denominator) and analogously for $q\to g_i q_j$. With the
$y_{ij}^{(D)}$ measure, eq.~(\ref{eq:yij}), a
branching that produces a soft quark, $E_j \ll E_i$, has the same
closeness as in the case of the gluon --- however this closeness is
now spurious because, in contrast to the gluon-emission case, there is
no divergence for $E_j \to 0$. The replacement of the $E_j$
denominator in the gluon-emission case, eq.~(\ref{eq:soft-col-mat}),
with $E_i$ in the `soft-quark' emission case,
eq.~(\ref{eq:soft-col-mat-g2qq}), suggests that the closeness measure
for soft $g\to q\bar q$ branching should become
$2\max(E_i^2,E_j^2)/Q^2 (1-\cos\theta_{ij})$. A similar argument holds
in the case of $q \to g_i q_j$ with $E_j \ll E_i$. Thus we should use
a distance measure that depends on the flavours of the particles being
considered:
\begin{equation}
  \label{eq:yij-flavour}
  y_{ij}^{(F)} = \frac{2(1-\cos\theta_{ij})}{Q^2} \times\left\{
    \begin{array}[c]{ll}
      \max(E_i^2, E_j^2)\,, & \quad\mbox{softer of $i,j$ is flavoured,}\\
      \min(E_i^2, E_j^2)\,, & \quad\mbox{softer of $i,j$ is flavourless,}
    \end{array}
  \right.
\end{equation}
where the softer of $i,j$ is the one with the smaller energy and where
we use the terms flavoured and flavourless rather than quark-like and
gluon-like so as to allow also for situations with diquarks or other
multi-flavoured objects.
With eq.~(\ref{eq:yij-flavour}) soft-quark `emission' leads to no
smaller a distance measure than non-soft quark emission, in accord
with the absence of a soft divergence for quark emission. Furthermore
if a quark is to recombine with a harder particle it will favour one
that is not too hard, in accord with the presence of $\max(E_i,E_j)$
in the denominator of eq.~(\ref{eq:soft-col-mat-g2qq}), which implies
that the harder the parent, the less likely it is that it will produce
a quark of a given softness.

With such a distance measure, for configurations as in
figure~\ref{fig:badjets} the soft $q$ and $\bar q$ will have similar
energies, $E_3\sim E_4 \ll Q$.  Thus $y_{13} \sim y_{14} \sim y_{23}
\sim y_{24} \sim 1$, whereas $y_{34} \sim E_3^2/Q^2 \ll 1$. So
independently of the precise (large) angles of the soft $\qq$ pair,
$3$ and $4$, it is that soft pair that will recombine first to give a
gluon-like pseudo-jet $g$. This will have $y_{1g} \sim y_{2g} \sim
E_2^2/Q^2$ and now the soft gluon pseudo-jet will recombine
with either $1$ or $2$ (which one depends on the angles) and the net
flavour of the hard particles will remain unchanged. Therefore, at
order $\as^2$, our new measure correctly eliminates the soft
flavour-changing divergence that exists for the plain Durham
algorithm.

Sometimes in the above algorithm a quark can be recombined with
another quark or with an antiquark of a different flavour. This can
happen for example if there are two large-angle $\qq$ pairs. As long
as the resulting `doubly-flavoured' object is treated in the same way
as a quark in the definition of $y_{ij}^{(F)}$, the algorithm will
remain infrared safe, because in the subsequent clustering steps there
will be a strong preference for recombining the multiply-flavoured
object with other objects of similar softness, until all soft
large-angle multiply-flavoured objects combine between themselves to
produce gluon-like objects (these then recombine normally with the
hard partons).

One may wish to avoid the appearance of multiply-flavour pseudo-jets
altogether, since they cannot be associated with QCD partons. This
can be achieved by vetoing any recombination that would lead to
a multiply-flavoured object, \ie by replacing step $2$ with
\begin{enumerate}
\item[2. (bland)] Find the specific $i$ and $j$ that correspond to the
  smallest $y_{ij}^{(F)}$ among those combinations of $i$ and $j$
  whose net flavour corresponds either to an (anti)quark or a gluon,
  and recombine them.
\end{enumerate}
We call this a `bland' variant of the jet-flavour algorithm, since
`excessively flavoured' clusterings are forbidden. We note that a
blandness requirement on clusterings has been discussed also in
\cite{CKKW} (though a simple `bland' Durham algorithm with the
standard $y^{(D)}_{ij}$ remains infrared unsafe).

An interesting question is that of how much freedom exists in the
definition of the distance measure for a flavour algorithm. Returning
to the analysis of fig.~\ref{fig:badjets} the main requirement for
infrared flavour safety is that the soft fermions $3$ and $4$ should
recombine between themselves before recombining with harder particles.
This property is maintained for the following \emph{class} of distance
measures,\footnote{We consider only those that reduce to the Durham
  algorithm for purely gluonic ensembles of particles.}
\begin{equation}
  \label{eq:yij-flavour-alpha}
  y_{ij}^{(F,\alpha)} = \frac{2(1-\cos\theta_{ij})}{Q^2} \times\left\{
    \begin{array}[c]{ll}
      [\min(E_i,E_j)]^{2-\alpha} [\max(E_i,E_j)]^{\alpha}\,, & 
       \quad\mbox{softer of $i,j$ is flavoured,}\\
      \min(E_i^2, E_j^2)\,, & \quad\mbox{softer of $i,j$ is flavourless,}
    \end{array}
  \right.
\end{equation}
where $\alpha$ is a continuous parameter in the range $0 <
 \alpha \le 2$ (so far we have implicitly discussed
 $\alpha=2$). Above, we stated the requirement that the distance
measure should not introduce `spurious' extra closeness for a
variation of the momenta that does not lead to any extra divergence.
Here though, for $\alpha < 2$ such a spurious extra closeness is
present. Infrared flavour safety is nevertheless preserved, because
the extra closeness is \emph{weaker} than that that arises in the case
of a divergence, i.e.\ for a soft gluon $j$, $y_{ij}$ vanishes as
$E_j^2$, whereas for a soft quark $j$ it only vanishes as
$E_j^{2-2\alpha}$.

Naively it would seem that $\alpha=2$ should give the best
identification of flavour. However there are situations where a hard
quark loses energy through multiple collinear gluon emission and thus
becomes a relatively soft quark. In principle there are no large
ratios between the quark energy and the softest of the harder gluons
it has emitted. However if that gluon is a bit harder than the quark,
a value of $\alpha<2$ can make it easier for them to recombine.
Accordingly below we shall examine both $\alpha=1$ and $\alpha = 2$.

\begin{figure}[htbp]
  \centering
  \includegraphics[width=0.65\textwidth]{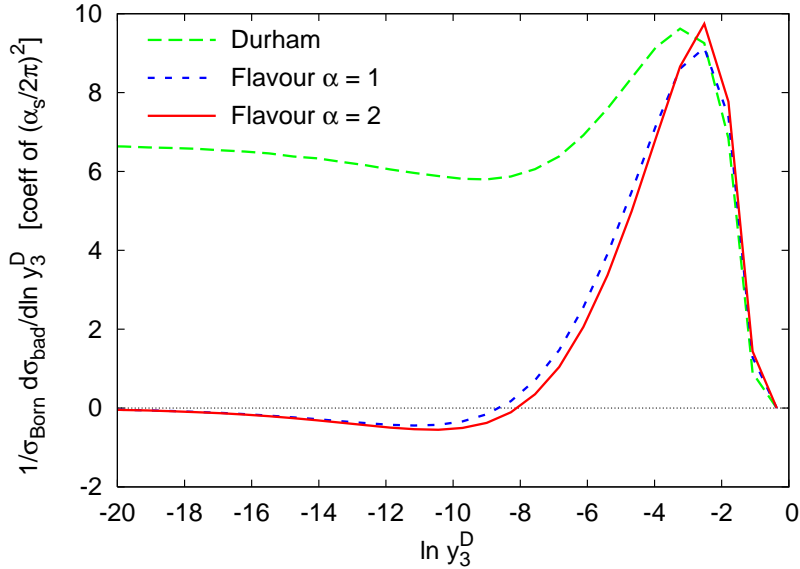}
  \caption{NLO differential cross section for $\ee \to q\bar q$
    events that after jet clustering have their flavour badly
    identified, \ie identified as consisting of two gluon jets (that
    is, each of zero net flavour) or two jets each of net flavour
    larger than 1; the coefficient of $(\as/2\pi)^2$, as generated
    with Event2 \cite{CataniSeymour}, is plotted as a function of the
    Durham $y_3$ three-jet resolution threshold; results are shown for
    the Durham and flavour algorithms (for two values of
    $\alpha$).}
  \label{fig:event2}
\end{figure}

A rigorous test of jet flavour algorithms can be obtained by
numerically investigating the infrared safety of the jet flavour in
fixed-order calculations. For example, one generates events $\ee \to
q\bar q$ together with higher orders and clusters them to two jets.
With a jet algorithm that provides a good reconstruction of the
flavour, one expects that each of the two jets should have net flavour
corresponding to an (anti)quark. Sometimes this does not happen ---
for example each of the two jets may have no net flavour, \ie be
gluon-like. This is legitimate in events in which there has been a
hard branching (there is not a unique clustering to two jets), but for
an infrared safe flavour jet algorithm, the probability of this
happening should vanish in the limit in which there are only soft and
collinear emissions.

To measure the hardness of a given event we use $y_3^D$, the threshold
value of the Durham jet-resolution below which the event is clustered
to three jets of more.\footnote{Any other global event-shape like
  variable that measures the departure from two jets could equally
  well have been used --- the only requirement is that for consistency
  in comparing the flavour behaviour of different jet algorithms one
  always use a
  common measure for determining the hardness of the event.} %
Figure~\ref{fig:event2} shows the differential cross section at next-to-leading
order (NLO, order $\as^2$) for producing events in which the flavour
of the two jets is badly identified. It has been obtained with
Event2~\cite{CataniSeymour}, to our knowledge the only NLO code that
provides information on the flavour of the final-state
partons.\footnote{In the default version of Event2 there were
  subtraction terms that had contributions from final states with
  different flavours --- for our studies here we split those
  subtraction terms so that each one corresponded to a unique set of
  final-state flavours.} %
One sees that for the Durham algorithm the differential cross section
for events 
whose jet flavour does not corresponds to $q\bar q$ goes to a constant
as $\ln y_3^D$ goes to $-\infty$.  This is the sign of the infrared unsafety of
flavour identification in the Durham jet algorithm. In contrast, in
our flavour algorithms (for both values of $\alpha$) the corresponding
cross section vanishes for $\ln y_3^D\to-\infty$. Detailed examination of the
events with badly identified flavour at small $y_3^D$ reveals that one
of the (anti)quarks has lost nearly all of its energy to a hard
splitting and goes into the same hemisphere as the other quark, \ie
identification of the event as consisting of two gluon jets is
actually legitimate. Such configurations appear at order $\as$ where
their cross section is $d\sigma_1^\mathrm{bad}/d\ln y_3 \sim \as
\sqrt{y_3^D}$.  At NLO, Sudakov suppression of an extra soft gluon
leads to a contribution %
\begin{equation}
  \label{eq:y3bad}
  \frac{d\sigma_2^{\mathrm{bad}}}{d\ln y_3} \simeq
  -\frac{\as}{2\pi} \left(\frac{C_A}{2}+\frac{C_F}{4}\right) \ln^2
  y_3^D \cdot \frac{d\sigma_1^\mathrm{bad}}{d\ln
    y_3} \sim \as^2 \sqrt{y_3^D} \ln^2 y_3^D\,,
\end{equation}
which is found to be consistent with the observed numerical results,
thus confirming the interpretation given above for the origin of the
small fraction of $gg$-like events at small $y_3^D$.

Given that one of the possible applications of jet flavour algorithms
is in the merging of matrix-element and parton-shower calculations, we
also wish to examine how flavour algorithms behave for Monte-Carlo
generated parton-level ensembles of quarks and gluons. This is
interesting for various other reasons too: Monte Carlo generators
produce multiple soft and collinear gluon emissions and $g\to q\bar q$
splittings, so they are more likely to `stress-test' a jet flavour
algorithm; also we can study a much wider variety of processes with
them --- for example one can simulate a fake $\ee \to gg$ to examine
jet flavour algorithms in a simple gluonic context; one can also easily
use them for studies of hadron-hadron events (next section) where
currently none of the NLO programs gives direct access to information
on the flavour of the outgoing partons.

While Monte Carlo event generators provide considerable flexibility,
it can be difficult to interpret their results. For example infrared
unsafety of the flavour in fixed-order programs manifests itself as a
non-vanishing probability of misidentification of the flavour as
$y_3^D \to 0$. With an event generator one is instead likely to see
this probability vanishing with an anomalous dimension, \eg
$(y_3^D)^{c \as}$ where $c$ is some coefficient (assuming, for the
purpose of the discussion, fixed coupling).

For an infrared safe jet flavour one expects that for the clustered
jets to have a different flavour from the Born channel there should
have been a hard branching, as in the discussion above for the NLO
$\ee$ calculation. This would lead to flavour misidentification
vanishing as $(y_3^D)^{d}$ where $d$ is some pure number (above,
$d=1/2$). This too may however be modified by an anomalous dimension,
becoming for example $(y_3^D)^{d + e \as}$ where $e$ is some further
pure number.\footnote{One kind of diagram that leads to flavour
  misidentification is that in which a hard quark loses
  most of its momentum by repeated gluon emission, and ends up in the
  opposite jet. This is similar to non-singlet small-$x$ quark
  production in parton distribution functions, known to be enhanced by
  an all-order double logarithmic series \cite{Bartels:1995iu}. Such
  a series might also appear in the jet-flavour case, leading to a more
  complex modification of the naive $(y_3^D)^{d}$ behaviour than
  stated in the main text.}

In the presence of anomalous dimensions it is difficult to establish
from Monte Carlo events exactly which functional form one is seeing.
Yet another complication is that Monte Carlo event generators often do
not contain the full structure of soft large-angle divergences, so
that in any case the anomalous dimensions observed may not correspond
to the true ones.

\begin{figure}[htbp]
  \centering
  \includegraphics[width=0.48\textwidth]{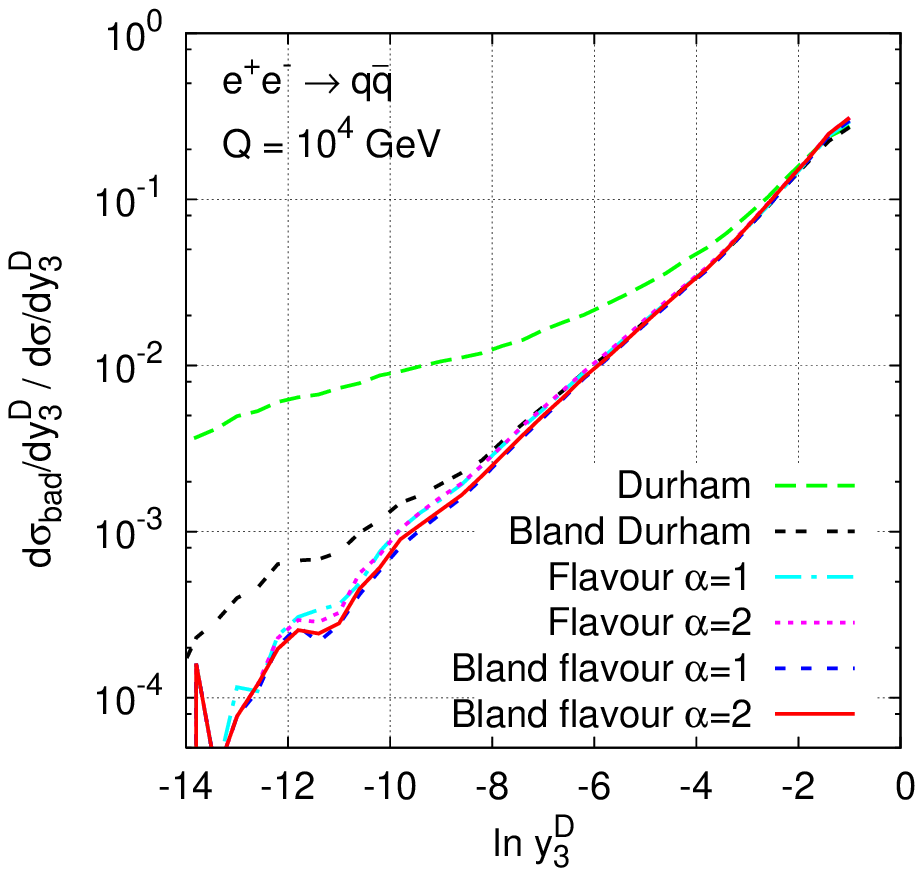}\hfill
  \includegraphics[width=0.48\textwidth]{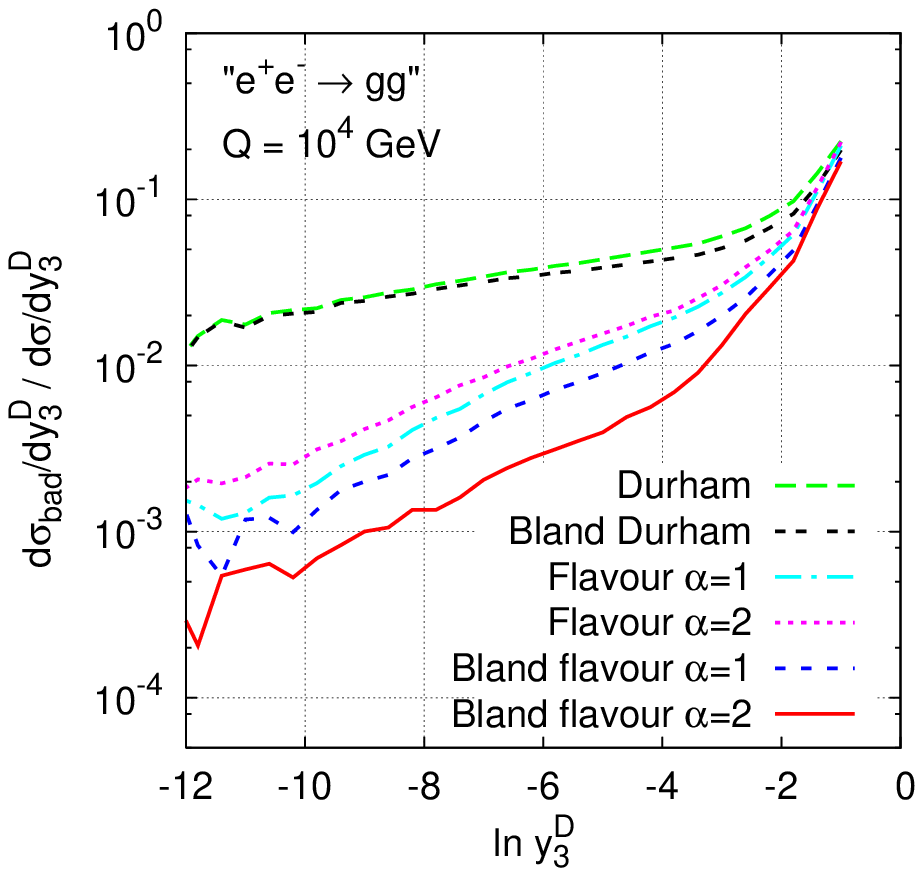}
  \caption{Fraction of events (generated by Herwig \cite{Herwig} at
    parton level) whose flavour is badly identified by various jet
    algorithms, shown as a function of the Durham $y_3^D$ jet
    resolution threshold; a large value of $Q$ has been chosen for
    illustrational purposes, so as to provide a correspondingly large
    range in $y_3^D$; the left-hand plot shows results for $\ee \to
    q\bar q$, while the right-hand plot shows fake ``$\ee \to gg$''
    process as generated by Herwig (code=107).}
  \label{fig:herwig-ee2xx}
\end{figure}

Despite these complications, for an infrared safe jet flavour
algorithm one expects flavour misidentification to vanish visibly
faster as $y_3^D \to 0$ than for the infrared unsafe case. This signal
can be made clearer by going to large $Q$ so as to have access to a
large range in $y_3^D$ (note though that a large value of $Q$ also
`stresses' the jet flavour algorithm, since it increases the phase
space for extra soft $q\bar q$ production).
Figure~\ref{fig:herwig-ee2xx} (left) shows the fraction of events, for
each $y_3^D$ value, where the flavour has been misidentified in
various jet algorithms. It has been generated for $Q=10^4$~GeV, using
Herwig \cite{Herwig} (chosen because it provides default access also
to a fake $\ee \to gg$ reaction, code $107$).

One sees clearly different $y_3^D$ dependences for the Durham versus
the flavour jet algorithms, with the flavour jet algorithm
misidentification vanishing considerably more rapidly (actually as
$\sqrt{y_3^D}$). Here all the flavour algorithms behave similarly.
Note also that the bland Durham algorithm works considerably better
than the plain Durham algorithm and only at very small $y_3^D$ values
does one see it doing worse than the flavour algorithms: for the bland
algorithm to generate a wrong-flavour event there must be a soft
$q\bar q$ pair of the same flavour as the hard $q\bar q$, and
additionally the directions of the soft $q \bar q$ must be such as to
lead to jets with net gluon flavour rather than diquark flavour.

This situation changes in the right-hand plot of
figure~\ref{fig:herwig-ee2xx}, where we consider fake $\ee \to gg$
events. Here the bland Durham algorithm behaves almost identically to
the normal Durham algorithm. This is expected, since a soft $q \bar q$
pair encounters no blandness problems when contaminating the flavour
of gluon jets. The flavour algorithms all work systematically better
than the Durham-based algorithms, clearly vanishing faster with
$y_3^D$. One sees differences in normalisation
between the different flavour algorithms and the blandness requirement
provides a non-negligible advantage, especially for $\alpha=2$. This
implies that the flavour misidentification involves more than one $q
\bar q $ pair. Nevertheless, the algorithm remains infrared safe even
for multiple soft or collinear $q \bar q$ pairs, as discussed
above\footnote{Note though that for a \emph{fixed} degree of softness, the
  presence of multiple $q \bar q$ pairs, spread densely in rapidity
  from large-angles all the way to the hard-fragmentation region can
  lead to a systematic worsening of the flavour identification.} (see
also the appendix for a more general outline of the discussion of IR
safety). 

%======================================================================
\section{Jet-flavour algorithms for hadron-hadron collisions}
\label{sec:flav-algor-hadr}

For hadron-hadron collisions (and DIS) the $k_t$ jet algorithm is
similar to that described in section~\ref{sec:k_t-flav-algor}, with a
few modifications in the definition of the distances
\cite{kthh,incl-kthh}. Given that there is no unique hard scale $Q$,
instead of examining dimensionless $y_{ij}$'s one looks at
dimensionful $d_{ij}$'s. These need to be invariant under longitudinal
boosts and the most widespread convention is to take
\begin{equation}
  \label{eq:dij}
    d_{ij} = \min(k_{ti}^2, k_{tj}^2) (\Delta \eta_{ij}^2 + \Delta
  \phi_{ij}^2)\,,\qquad
\end{equation}
where $\Delta\eta_{ij} = \eta_i - \eta_j$, $\Delta\phi_{ij} = \phi_i -
\phi_j$ and $k_{ti}$, $\eta_i$ and $\phi_i$ are respectively the
transverse momentum, rapidity and azimuth of particle $i$, with
respect to the beam. A particle $i$ can also recombine with the beam
and here too one needs a distance measure, usually taken to be
\begin{equation}
  \label{eq:diB}
    d_{iB} = k_{ti}^2\,.
\end{equation}
It is the smallest of the $d_{iB}$ and the $d_{ij}$ that determines
which recombination takes place. If it is $d_{iB}$ that is smallest at
a given step, then $i$ recombines  with the beam (or else gets called a
jet, in the ``inclusive'' version of the algorithm).

The modification of the $d_{ij}$ needed to obtain a flavour-safe jet
algorithm is directly analogous to that used for the $\ee$ algorithm:
\begin{equation}
  \label{eq:dij-flavour}
  d_{ij}^{(F)} = (\Delta \eta_{ij}^2 + \Delta \phi_{ij}^2) \times\left\{
    \begin{array}[c]{ll}
      \max(k_{ti}^2, k_{tj}^2)\,, & \quad\mbox{softer of $i,j$ is flavoured,}\\
      \min(k_{ti}^2, k_{tj}^2)\,, & \quad\mbox{softer of $i,j$ is flavourless,}
    \end{array}
  \right.
\end{equation}
where by `softer' we now mean that having lower $k_t$ and where
temporarily, for simplicity, we consider only the case
$\alpha=2$. 

It is less obvious how to modify the beam distance. The problem is
that $d_{iB}$ involves just a single scale, $k_{ti}^2$, and so there
is no ``minimum'' that one can replace with a ``maximum''. However one
could imagine that $d_{iB}$ is actually the minimum of $k_{ti}^2$ and
some transverse scale associated with the beam, $k_{tB}^2$, which has
never been explicitly needed so far because it was always larger than
any of the $k_{ti}^2$. The analogue of eq.~(\ref{eq:dij-flavour})
would then be to take 
\begin{equation}
  \label{eq:diB-flavour}
  d_{iB}^{(F)} = \left\{
    \begin{array}[c]{ll}
      \max(k_{ti}^2, k_{tB}^2)\,, & \quad\mbox{$i$ is flavoured,}\\
      \min(k_{ti}^2, k_{tB}^2)\,, & \quad\mbox{$i$ is flavourless.}
    \end{array}
  \right.
\end{equation}
The question that remains is how to define $k_{tB}$. 

A first issue is that we will want to identify the flavour of each of
the incoming beams. So whereas for the normal $k_t$ algorithm one
recombines particles with `the beams', here we will need to specify
\emph{which} of the two beams a particle recombines with. Therefore we
will need to define $k_{tB}$ for the beam moving towards positive
rapidities (right) and $k_{t\bar B}$ for the other beam.

In line with the DGLAP idea~\cite{DGLAP} of logarithmic ordering,
such that harder emissions are at successively larger angles with
respect to the beam that produced them, it makes sense for the beam
hardness to be a function of rapidity, $k_{tB}(\eta)$. In the
definition of $d_{iB}$, eq.~(\ref{eq:diB-flavour}), one would then use
$k_{tB}(\eta_i)$. For the right-moving (positive rapidity) beam, one
scale that appears naturally is (with $\Theta(0) \equiv 1/2$),
\begin{align}
  \label{eq:PtRightScale}
  P_{t,\rght}(\eta) &= \sum_i k_{ti} \Theta(\eta_i - \eta)\,,
\end{align}
\ie the beam scale should be at least as hard as all emissions that
have already occurred from that beam (\ie all emissions that are at
larger rapidity). Another scale that arises is 
\begin{equation}\label{eq:scale-alpha-left}
  P_{\alpha,\lft}(\eta) = \sum_i k_{ti} e^{\eta_i}\Theta(\eta -
  \eta_i)\,.
\end{equation}
When one performs a Sudakov decomposition of all momenta $k_i =
\alpha_i P + \beta_i \bar P + \vec k_{ti}$ ($P = (1,0,0,1)$ and $\bar
P = (1,0,0,-1)$), in the massless approximation, this scale is just
the sum of the $\alpha_i = k_{ti} e^{\eta_i}$ components of all
particles that are still to be emitted by this beam (\ie are at
smaller rapidity). It is equivalent to the light-cone momentum still
left in the beam. This scale depends on the reference frame, but can
be transformed into a boost invariant, local `transverse' hardness by
multiplying it by $e^{-\eta}$, giving\footnote{Another way of seeing
  how this scale arises naturally is to recall that in the
  non-longitudinally invariant version of the $k_t$ algorithm for
  DIS and hadron-hadron collisions~\cite{ktDIS}, the beam distance is
  $d_{iB} = 2 E_{i}^2 (1-\cos \theta_{iB}) $. Replacing $E_{i}$ with
  the effective beam energy $\frac12 P_{\alpha,\lft}$ (\ie taking the
  larger of $E_i$ and the effective beam energy) and taking the
  small-angle limit gives precisely $P_{t\alpha,\lft}^2$.}
\begin{equation}
  P_{t\alpha,\lft}(\eta) = \sum_i k_{ti} e^{\eta_i - \eta}\Theta(\eta -
  \eta_i)\,.\label{eq:scale-talpha-left}
\end{equation}
By adding the two measures, $P_{t,\rght}(\eta)$ and
$P_{t\alpha,\lft}(\eta)$ for the beam scale, one obtains an overall
beam hardness measure,
\begin{equation}
  \label{eq:ktB}
  k_{tB} (\eta) =
  % = P_{t,\rght}(\eta) + P_{t\alpha,\lft}(\eta)
  \sum_i k_{ti} \left( \Theta(\eta_i - \eta) +
    \Theta(\eta - \eta_i) e^{\eta_i - \eta}\right)\,,
\end{equation}
that takes into account both emissions that have already occurred at a
certain rapidity (in the picture of ordering of emissions) and those
that will occur further on. Similarly one defines a scale for the
other beam
\begin{equation}
  \label{eq:ktBbar}
  k_{t\bar B} (\eta) = 
  %P_{t,\lft}(\eta) + P_{t\beta,\rght}(\eta) = 
  \sum_i k_{ti} \left( \Theta(\eta - \eta_i) +
    \Theta(\eta_i - \eta) e^{\eta - \eta_i}\right)\,.
\end{equation}
In the same way that one updates the $d_{ij}$ and $d_{iB}$ after each
clustering, one should update also the $k_{tB}$ and $k_{t\bar B}$.

\begin{figure}[htbp]
  \centering
  \includegraphics[width=0.6\textwidth]{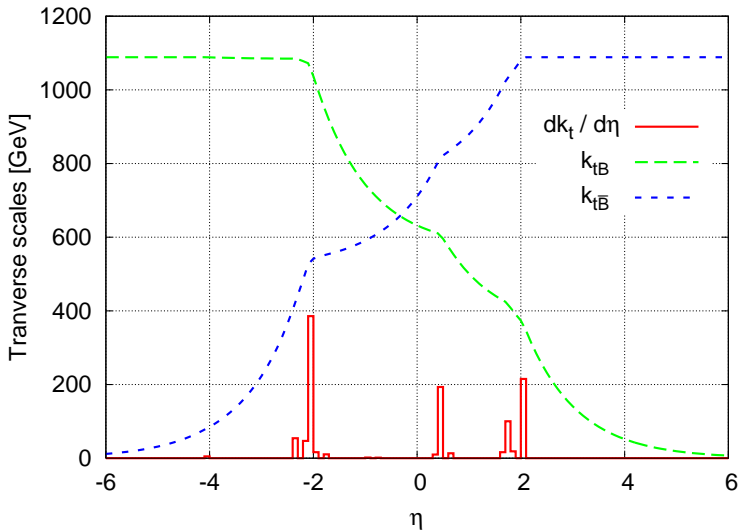}
  \caption{Plot of $k_{tB}$ and $k_{t\bar B}$ for a multi-jet
    parton-level LHC event, generated by Herwig; also shown is the
    histogram of the rapidity distribution of transverse momenta.}
  \label{fig:ktb-event}
\end{figure}

To illustrate the properties of $k_{tB}$ and $k_{t\bar B}$,
fig.~\ref{fig:ktb-event} shows these two quantities for a typical
multi-jet LHC event (represented as a histogram of total transverse
momentum per bin of rapidity). 
Towards positive rapidities, $k_{tB}(\eta)$ decreases as $e^{-\eta}$,
while $k_{t\bar B}(\eta)$ approaches a constant, so that as is
natural, positive-rapidity particles combine with $B$, while negative
rapidity particles combine with $\bar B$. At the point where $k_{tB}$
and $k_{t\bar B}$ cross, they are of the same order of magnitude as
the total transverse momentum in the event, \ie its overall hardness.
Note also that $k_{tB}(\eta)$ and $k_{t\bar B}(\eta)$ are always at
least as hard as the hardest emission at rapidity $\eta$.

Let us now summarise the jet flavour algorithm for hadron-hadron
collisions:%
\begin{enumerate}
\item Introduce a distance measure $d_{ij}^{(F)}$ between every pair
  of partons $i$, $j$:
  \begin{equation}
    \label{eq:dij-flavour-alpha}
    d_{ij}^{(F,\alpha)} = (\Delta \eta_{ij}^2 + \Delta \phi_{ij}^2)
    \times\left\{ 
      \begin{array}[c]{ll}
        \max(k_{ti}, k_{tj})^\alpha \min(k_{ti}, k_{tj})^{2-\alpha}\,,
        & \quad\mbox{softer of $i,j$ is flavoured,}\\
        \min(k_{ti}^2, k_{tj}^2)\,, & \quad
        \mbox{softer of $i,j$ is flavourless,}
      \end{array}
    \right.
  \end{equation}
  as well as distances to the two beams,
  \begin{equation}
    \label{eq:diB-flavour-alpha}
    d_{iB}^{(F,\alpha)} = \left\{
      \begin{array}[c]{ll}
        \max(k_{ti}, k_{tB}(\eta_i))^\alpha 
        \min(k_{ti}, k_{tB}(\eta_i))^{2-\alpha}
        \,, & \quad\mbox{$i$ is flavoured,}\\
        \min(k_{ti}^2, k_{tB}^2(\eta_i))\,, & \quad\mbox{$i$ is flavourless,}
      \end{array}
    \right.
  \end{equation}
  and an analogous definition of $d_{i\bar B}^{(F,\alpha)}$ involving
  $k_{t \bar B}(\eta_i)$ instead of $k_{tB}(\eta_i)$ (both defined as
  in eqs.~(\ref{eq:ktB}) and (\ref{eq:ktBbar})).\footnote{The beam
    distances in eqs.~(\ref{eq:ktB}) and (\ref{eq:ktBbar}) have been
    constructed by considering situations with just massless partons.
    However, their definition can be extended to cases with massive
    particles in the final state by replacing $k_{ti}$ with
    $\sqrt{k_{ti}^2+m_i^2}$. Notice that any heavy non-QCD particles
    should also be included in the sums~(\ref{eq:ktB}) and
    (\ref{eq:ktBbar}), even if they do not enter the clustering.
    In DIS, in the Breit frame, $k_{tB}(\eta)$ should include an
    additional contribution related to the virtual photon, given by
    $Q(\Theta(\eta)e^{-\eta} + \Theta(-\eta))$, while $k_{t\bar
      B}(\eta)$ should have an additional contribution $Q(\Theta(\eta) +
    \Theta(-\eta)e^{\eta})$, where $Q$ is the photon virtuality.
  } %
  As in section~\ref{sec:k_t-flav-algor} we have introduced a class of
  measures, parametrised by $0<\alpha\le 2$.
  
\item Identify the smallest of the distance measures. If it is a
  $d_{ij}^{(F,\alpha)}$, recombine $i$ and $j$; if it is a
  $d_{iB}^{(F,\alpha)}$ ($d_{i\bar B}^{(F,\alpha)}$) declare $i$ to be
  part of beam $B$ ($\bar B$) and eliminate $i$; in the case where the
  $d_{iB}^{(F,\alpha)}$ and $d_{i\bar B}^{(F,\alpha)}$ are equal
  (which will occur if $i$ is a gluon), recombine with the beam that
  has the smaller $k_{tB}(\eta_i)$, $k_{t\bar B}(\eta_i)$.

\item Repeat the procedure until all the distances are larger than
  some $d_{cut}$, or, alternatively, until one reaches a predetermined
  number of jets.\footnote{Yet another possibility is to introduce
    separate measures for the ordering of recombinations and for the
    point where recombination comes to a stop, as in the Cambridge and
    Aachen algorithms~\cite{CamAachen}.}$^,$%
  \footnote{In light of recent work that relates the $k_t$ algorithm to
  a geometrical nearest neighbour problem \cite{FastJet} to reduce its
  computational complexity to $N \ln N$, it is worth commenting that
  the simultaneous use here of both $\min(k_{ti}^2,k_{tj}^2)$ and
  $\max(k_{ti}^2,k_{tj}^2)$ invalidates the Lemma of \cite{FastJet}
  that was central in making the connection with a nearest neighbour
  problem.  It is therefore not clear whether it would be possible to
  write the flavour algorithm such that its complexity goes as $N \ln
  N$. The implementation that we use has a complexity that scales
  roughly as $N^2$.}

\end{enumerate}

In the `bland' variant of the algorithm one considers only those
$d_{ij}$ for which the product of the recombination would have at most
one flavour. Similarly one considers only a subset of the $d_{iB}$ ---
in this case the blandness requirement is imposed on the flavour of
the parton entering the hard interaction, or equivalently on the
difference between the flavour of the incoming hadron and the flavour
contained in the outgoing beam jet.

The infrared safety of this algorithm follows from the same arguments
that were used in the $\ee$ context. The beam scales simply ensure
that $q\bar q$ pairs that are soft but separated by $\Delta\eta^2 +
\Delta\phi^2 > 1$ recombine with each other before recombining with
the beam. This eliminates the potentially dangerous situation that
would otherwise occur, in which first the $q$ recombines with one beam
and then the $\bar q$ recombines with the other beam. Therefore it is
not just the flavours of the outgoing jets that are infrared and
collinear safe, but also those of the incoming beam jets (the
determination the beam-jet flavours of course also requires knowledge
of the incoming parton flavours).

A concrete demonstration of the infrared safety of the hadron-hadron
algorithms, analogous to figure~\ref{fig:event2} for $\ee$, is not
possible with currently available tools, because none of the
higher-order NLO jet codes \cite{Kilgore:1996sq,NLOJET} provide direct
access to information about final-state flavour. Even if they did,
there would be an additional complication compared to $\ee$. In $\ee$
at Born level, there is only one flavour channel, \ie $\ee \to q\bar
q$. Therefore one could identify flavour infrared unsafety by
examining, for example, the 3-jet NLO cross section for jets
classified as $gg$. In hadron-hadron collisions all flavour channels
are present at Born order, therefore to verify the infrared safety of,
say, the $gg \to gg $ channel one must supplement the NLO 2+3 jet
calculation with the $gg\to gg$ Born contribution and its two loop
corrections, \ie one must carry out a NNLO 2+2 jet calculation, which
is beyond today's technology.  Fortunately an alternative method
exists for verifying the IR safety of flavour identification using
just a NLO 2+3 jet
calculation, namely by examining the cross section for
doubly-flavoured jets, since these do not appear at Born level, but
are infrared unsafe in the plain $k_t$ algorithm. We hope that flavour
information will soon become available in $2+3$ jet NLO codes, making
it possible to demonstrate this explicitly.

In the absence of any way of obtaining a fixed-order illustration of
the infrared safety of the flavour algorithms, we resort to
investigations of reconstruction of the flavour in parton-level Monte
Carlo events. This is achieved by comparing, event-by-event, the
flavours in the hard $2\to 2$ partonic scattering with those of the
beam and outgoing jets after clustering of the event to $2+2$
jets. 
Since the normal $k_t$ algorithm does not usually distinguish between
the two beams, we extend it (both normal and bland variants) such that
a particle destined to recombine with `the beams' is assigned to that
with the smaller of $k_{tB} (\eta_i)$ and $k_{t\bar B} (\eta_i)$.
 
\begin{figure}[htbp]
  \centering
  \includegraphics[width=0.47\textwidth]{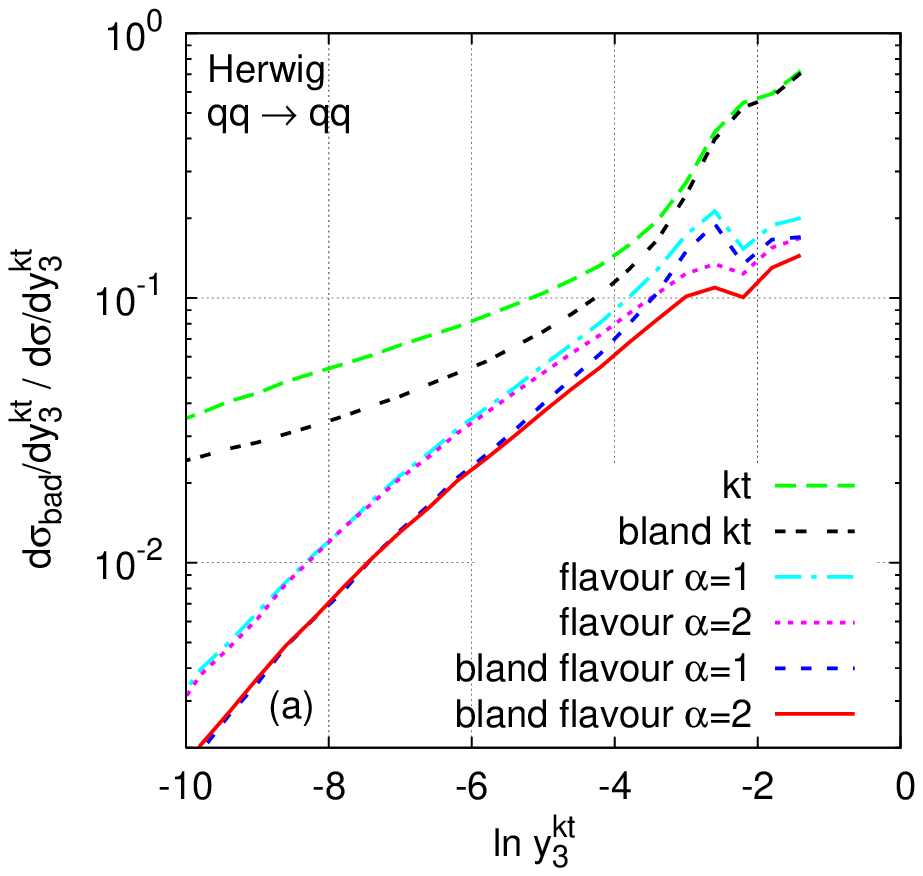}\hfill
  \includegraphics[width=0.47\textwidth]{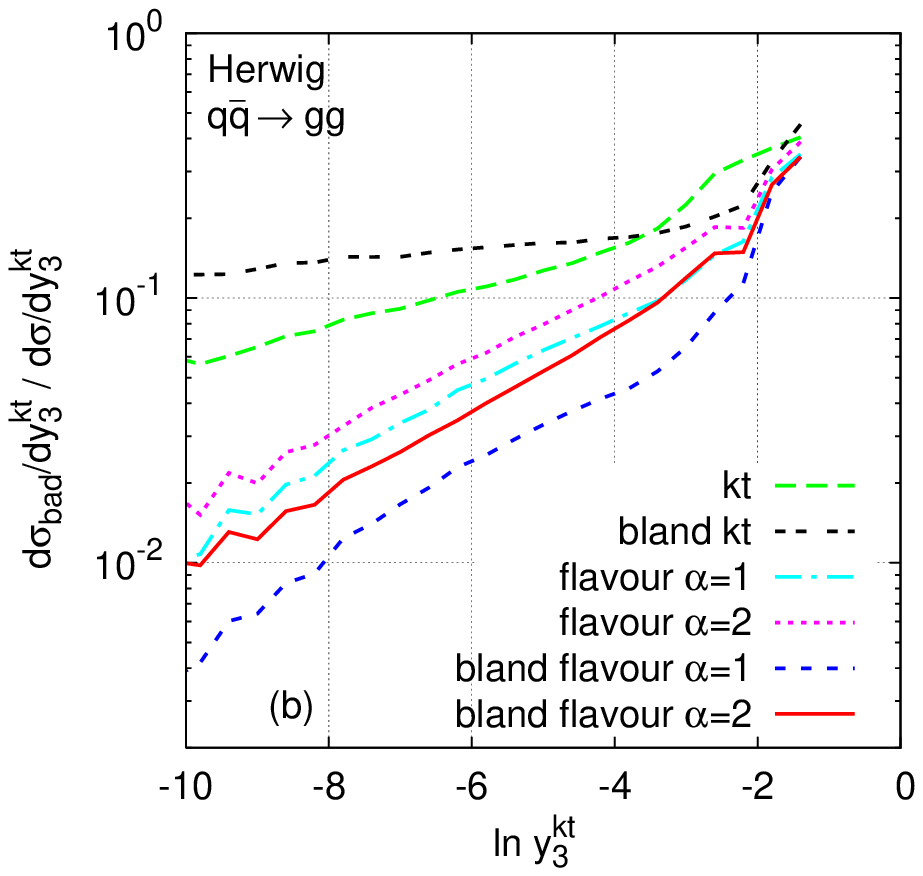}\\
  \includegraphics[width=0.47\textwidth]{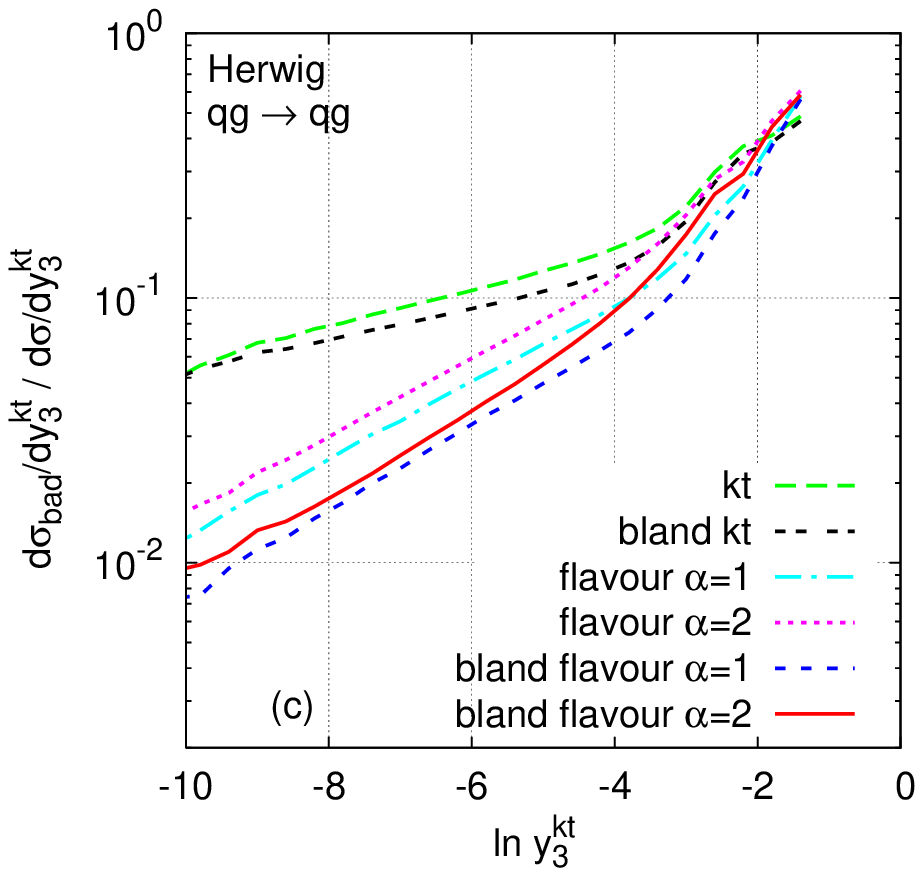}\hfill
  \includegraphics[width=0.47\textwidth]{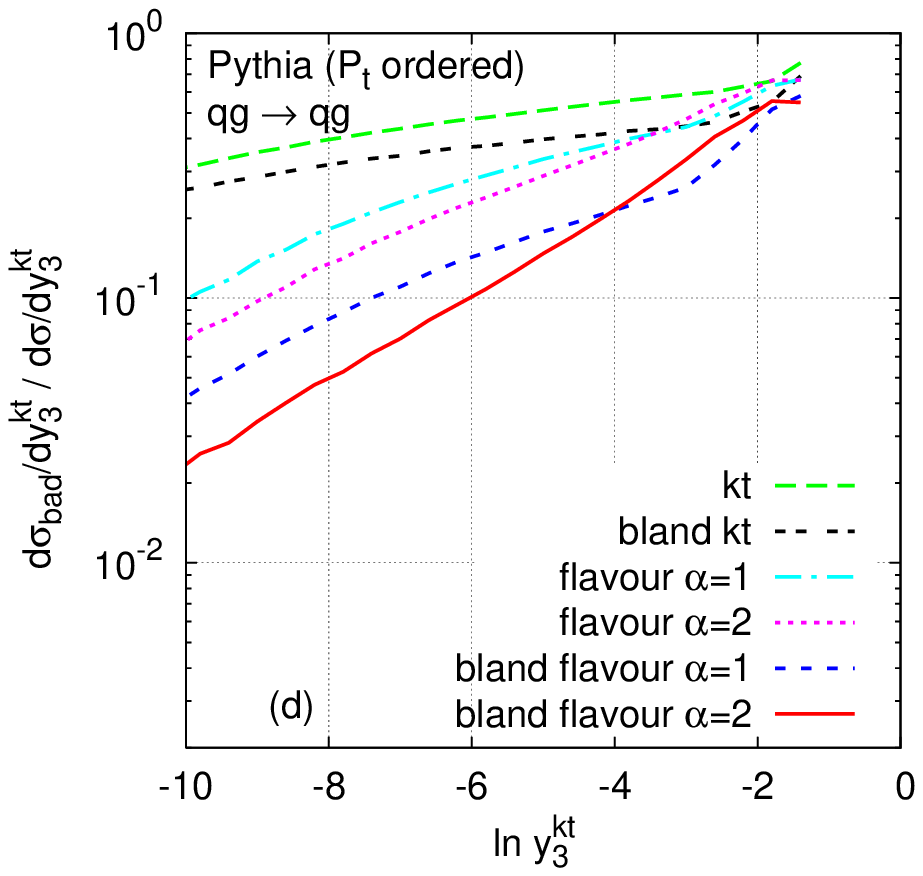}\\
  \caption{The proportion of Monte Carlo events in which the flavour
    of one or more incoming or outgoing reconstructed parton-level
    jets differs from the flavour in the corresponding parton in the
    original hard event; shown as a function of $\ln y_{3}^{kt}$ for three
    channels (in the case of $qg \to qg$ for both Herwig \cite{Herwig}
    and a recently developed parton shower algorithm in Pythia
    \cite{PtOrderedPythia}); LHC kinematics are used and the events
    selected are those where the hardest $k_t$-algorithm jet has a
    transverse momentum greater than $1\TeV$ and where the two hardest
    jets have $|\eta| < 1$. The range of most common values of $y_3^{kt}$
    depends on the subprocess but is typically roughly $-8 \lesssim
    \ln y_3^{kt} \lesssim -3$.}
  \label{fig:hhXX2YY}
\end{figure}

The proportion of events where the original and reconstructed $2\to2$
flavours do not match is shown in fig.~\ref{fig:hhXX2YY}, as a
function of $y_3^{kt} = d_3^{kt}/(E_{t1}+E_{t2})^2$.  Here $d_3^{kt}$
is the threshold value of $d_{cut}$ below which the event is clustered
to 3 or more jets in the standard exclusive longitudinally-invariant
$k_t$ algorithm \cite{kthh}; $E_{t1}$ and $E_{t2}$ are the transverse
energies of the two last jets to be recombined with the beam if there
is no $d_{cut}$ \cite{hhresum} (equivalently the two hardest jets when
running the inclusive $k_t$ algorithm \cite{incl-kthh}). We consider
simulated LHC events and require the hardest jet to have a transverse
energy larger than $1\TeV$ and the two hardest jets to have $|\eta|<1$.

Three representative channels, $qq \to qq$ (including $q\bar q \to
q\bar q$), $q\bar q \to gg$ and $qg \to qg$ are shown in
fig.~\ref{fig:hhXX2YY}, as obtained with Herwig \cite{Herwig}.  The
standard parton showering in Pythia \cite{Pythia} gives similar
results (with a slightly higher normalisation). We also illustrate the
$qg \to qg$ channel using the recently developed transverse momentum
ordered shower in Pythia \cite{PtOrderedPythia}. In all cases one sees
that the rate of flavour misidentification falls significantly more
rapidly towards small $y_3^{kt}$ for the flavour algorithms than for
the normal $k_t$ algorithm or its bland variant.\footnote{It is
  interesting to note that the bland $k_t$ algorithm sometimes behaves
  worse than the normal $k_t$ algorithm (\eg for $q\bar q \to gg$). To
  see why this happens, consider a beam corresponding to incoming $u$
  flavour, together with a soft collinear $u \bar u$ pair. In the
  normal $k_t$ algorithm, the $u$ and $\bar u$ can recombine with the
  beam in any order. In the bland variant the $\bar u$ is prevented
  from recombining first (because the parton entering the reaction
  would then implicitly have $uu$ flavour) and if it has the lower
  $k_{t}^2$ it will instead try to recombine with the other (wrong)
  beam. Therefore the bland algorithm actually has an extra source of
  infrared-collinear flavour unsafety relative to the plain
  $k_t$ algorithm.} %
This is similar to what was observed
for $\ee$ in section~\ref{sec:k_t-flav-algor}, and is a sign of the
infrared safety of the flavour algorithms.

One notes that for all algorithms the fall-off is less rapid in the
hadron-hadron case than in $\ee$. This is natural given the
increased number of jets and therefore of sources of radiation which
can lead to extra flavour in the final state. Another difference
compared to $\ee$ is that now the $\alpha=1$ flavour algorithms
sometimes fare better than the $\alpha=2$ case. This is not systematic
and also depends on the Monte Carlo program used to generate events
(compare figs.~\ref{fig:hhXX2YY}c and \ref{fig:hhXX2YY}d). The overall
normalisation of the curves also depends on the Monte Carlo program
used and one sees that Pythia with transverse-momentum ordered showers
produces parton-level final states in which it is systematically
harder to cluster back to the original flavour.

%======================================================================
\section{Outlook}
\label{sec:conclusions}

We have shown in this article that it is possible to define
parton-level jets in a manner that ensures that their flavour is
infrared safe. The key ingredient in doing this was a modification of
the $k_t$ distance measure, inspired by the different structures of
divergences that appear in quark production and gluon production. In
the case of hadron-hadron collisions it was also necessary to
introduce the concept of a hardness associated with the beam at any
given point in rapidity. Where possible, explicit NLO verifications
confirm the infrared safety of the new `flavour' jet
algorithms. Parton-level Monte Carlo studies also indicate a
significant improvement in the identification of flavour relative to
the $k_t$ algorithm.

To make use of our new algorithms to accurately study jet flavour, it
is necessary to have access to information about the flavour of
final-state partons in NLO jet codes. Currently however, most NLO jet
codes have been designed assuming that the user has no need for
information about final-state parton flavour (an exception is Event2
\cite{CataniSeymour}). In light of the developments presented here, we look
forward to flavour information being made available in the future (\eg
\cite{NagyInProgress}).

Our original motivation for studying the problem of jet flavour was
the need to accurately combine resummed predictions for
hadron-collider dijet event shapes \cite{caesar,hhresum} with
corresponding fixed-order predictions \cite{Kilgore:1996sq,NLOJET}.
Another simple flavour-related study would be the investigation of how
the relative fractions of quark and gluon jets at hadron colliders are
modified by NLO corrections and how they vary with jet transverse
momentum. Apart from its intrinsic interest, such information could be
of relevance also to the tuning of Monte Carlo event generators and
studies of hadron multiplicities in jets, both of which are sensitive
to the proportions of quark and gluon jets.

One drawback of the algorithms presented here is that, when
considering light flavours, they can only be applied to partonic
and not hadronic events. This is because at each recombination
they require knowledge of which objects are flavoured (quark-like) rather than
flavourless (gluon-like) and that information is not present in
hadronic final states..
It would be interesting
to find a jet algorithm based purely on particle momenta, that
nevertheless provides a good infrared-safe determination of the
flavour at parton level. It is not clear to what extent this is
possible.\footnote{A candidate for a jet flavour algorithm that does
  not use flavour information during the recombination sequence might
  be the JADE algorithm \cite{jade}. The flavour of its jets can be
  shown to be infrared safe in $\ee$ at $\order{\as^2}$.  However it
  has numerous other drawbacks, which we suspect are part of the
  reason why in Monte Carlo studies we find that its flavour
  identification properties are no better than those of the
  (flavour) infrared unsafe Durham algorithm.} %

There is nevertheless one hadron-level context in which this article's
flavour algorithms could be used directly, that is for heavy-quark
jets~\cite{HQJets}.  Currently a heavy-quark jet is defined as a jet
containing one or more heavy quarks (or heavy-quark hadrons). The
fraction of jets of transverse energy $E_T$ containing a heavy quark
of mass $m_Q$ is enhanced by terms $\as^n \ln^{2n-1} E_T/m_Q$ for $E_T
\gg m_Q$, due to the large multiplicity $\sim \as^n \ln^{2n} E_T/m_Q$
of gluons above scale $m_Q$, combined with the possibility that they
split collinearly $g \to Q\bar Q$, responsible for a further factor
$\as \ln E_T/m_Q$ \cite{Mueller:1985zp,Mangano:1992qq,Seymour:1994ca}.
Therefore, at high $E_T$ the majority of so-called heavy-quark jets
are not jets induced by a heavy quark, but rather jets in which a
heavy quark has appeared from the internal branching in the jet. This
implies that the current definition of heavy-quark jet will lead to
large QCD backgrounds in searches for new particles which aim to tag
an `intrinsic' heavy quark jet among the decay products of the new
particle.

An alternative approach to the study of heavy quark jets would be to
consider the \emph{net} heavy flavour of jets,\footnote{A study that
  goes partially in this direction is the recent investigation of
  angular correlations between $b\bar b$ pairs \cite{Acosta:2004nj}.}
\ie the number of heavy quark hadrons minus heavy antiquark hadrons
in a given jet.\footnote{In an event with just two heavy hadrons one
  need not know which one is quark-like and which antiquark-like ---
  it suffices to know that if combined they give zero net heavy
  flavour.}  With the cone or $k_t$ algorithms such a definition would
eliminate nearly all the final-state logarithmically enhanced terms,
leaving just $\as^n \ln^{n-1} E_T/m_Q$ contributions (involving a
final-state BFKL-type resummation
\cite{Marchesini:2003nh,Marchesini:2004ne}).  These remaining terms
come from the same diagrams that led to the infrared unsafety of light
flavour of a jet. They can therefore be eliminated altogether by
applying our flavour jet algorithm with the minor modification that
every occurrence of ``flavour'' is to be replaced with ``heavy
flavour''. In this way it becomes possible to give meaning to a
concept of intrinsic heavy flavour, \ie heavy flavour that originates
exclusively from the heavy-flavour component of parton distribution
functions, from hard QCD flavour ``creation'' (\eg $gg\to Q\bar Q$) and
from the decay of other
heavier particles.  We look forward to future phenomenological
investigation of this concept.

%======================================================================
\subsection*{Acknowledgements}

All three of us are grateful to the Milano Bicocca theoretical physics
group for hospitality during the initial stages of this work. GPS also
wishes to thank Fermilab, the Cavendish Laboratory and CERN for
hospitality. We thank Yuri Dokshitzer for several discussions and
suggestions, Matteo Cacciari and Giuseppe Marchesini for comments,
Bryan Webber for bringing to our attention the original use of the
`bland Durham' algorithm in CKKW matching, Michael Seymour for
assistance with Event2, Peter Skands for help with Pythia, and William
Kilgore and Zoltan Nagy for information about the accessibility of
flavour information in hadron-collider NLO jet codes.

This work was supported in part by grant ANR-05-JCJC-0046-01 from the
French Agence Nationale de la Recherche.

\appendix
%======================================================================
\section*{Appendix}

The arguments for the infrared safety of our jet flavour algorithms,
as discussed in section~\ref{sec:k_t-flav-algor}, applied only to the
case of one or two extra soft $q \bar q$ pairs. Here we give an
outline of a general all-order discussion of infrared and collinear
safety of the flavour. It will be framed in the context of $\ee$
collisions, and then in closing we will briefly mention hadronic
collisions.

For a general discussion of the infrared and collinear safety of
flavour one needs to examine all divergent cases in which flavour is
either produced or moved from one part of the event to another. 
Production of flavour arises from gluon splitting. This has just a
  collinear divergence; additionally the gluon itself has soft and
  collinear divergences with respect to other quarks and gluons.
Flavour can `move' during the branching process when a quark recoils
  due to emission of a gluon of similar hardness to the quark. This
  has no divergences, but there may be divergences associated with the
  original production of the quark itself.
Flavour can also move during the jet-clustering procedure whenever a
  quark recombines with a parton that is not collinear to it and whose
  momentum is of the same order of magnitude as (or larger than) the
  quark.

Let us first consider flavour production by collinear splitting of a
gluon. The Durham algorithm always recombines collinear particles into
the same jet. Since in $g\to q\bar q$ splitting there is no soft
divergence, the $q$ and $\bar q$ have commensurate hardnesses.
Therefore the `flavour' distance measure eq.~(\ref{eq:yij-flavour}) is
of the same order of magnitude as the Durham distance measure and so
the $q\bar q$ from a collinear splitting of a gluon will end up in the
same jet also in the flavour algorithm, leaving the jet flavour
unchanged as is required for IRC safety of the flavour.

Next we consider non-collinear splitting of a gluon into $q \bar q$.
This has divergences when the original gluon is collinear to some
other parton and/or soft. If the gluon itself is collinear to some
other parton $a$, angle $\theta_{ag} \ll 1$, then the gluon splitting
to $q \bar q$ is strongly suppressed unless $\theta_{q\bar q} \sim
\theta_{ag}$, \ie non-collinear splitting is not possible from a gluon
that is collinear to some other parton. This is the basis of the
widely used angular ordering approximation. Therefore a $q \bar q$
produced from a collinear (and optionally soft) gluon will always
recombine, in the flavour algorithm as in the Durham algorithm,
ensuring the safety of the flavour of any resulting jet.

This leaves the case discussed already in the main text, in which a
large-angle $q \bar q$ pair is produced from a large-angle soft
gluon. We have already presented the arguments that explain the IR
unsafety of the Durham algorithm in this case and the IR safety of the
flavour algorithms.

In generalising the analysis to higher orders one needs also to
examine potential `motion' of the soft large-angle $q$ and $\bar q$.
It will be useful to introduce the compact notation $y_{1\{2\ldots
  n\}}$ for the set of distance measures
$y_{12},y_{13},\ldots,y_{1n}$.

\begin{figure}[htbp]
  \centering
  \begin{minipage}[b]{0.33\textwidth}\centering
    \includegraphics{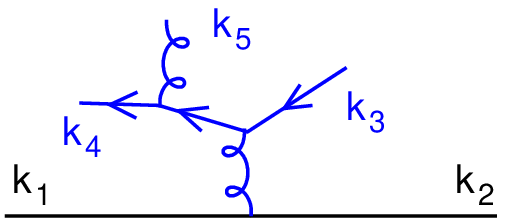}
  \end{minipage}\;
  \begin{minipage}[b]{0.33\textwidth}\centering
    \includegraphics{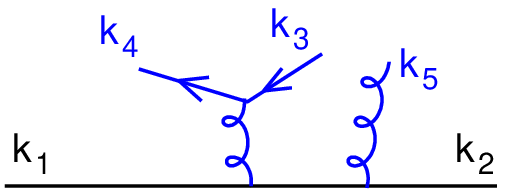}
  \end{minipage}
  \caption{Configurations in which flavour `moves' during branching
    and clustering, discussed in the text with regards to infrared and
    collinear safety.}
  \label{fig:badjets-app}
\end{figure}

Firstly the quark (or anti-quark) can itself emit a large-angle gluon
of similar softness ($k_5$), fig.~\ref{fig:badjets-app} (left) . This
will change the direction of the quark ($k_4$). In the Durham
algorithm, each of the $y_{\{12\}\{345\}}$ is of the order of the soft
gluon $k_t^2/Q^2$, and the recombination sequence depends
significantly on the angles. In particular the emission of $k_5$ from
the quark may have moved it further away from the antiquark making it
more likely that the soft $q\bar q$ end up in different jets. In
contrast, in the flavour algorithm $y_{\{12\}\{34\}}$ are of order
$1$, whereas $y_{\{1234\}5}$ and $y_{34}$ are of order of the soft
gluon $k_t^2/Q^2$. Therefore $3$, $4$ and $5$ will all recombine
together first, or $5$ will recombine with the hard jets and then $3$
and $4$ will recombine together. In both cases the flavour of the soft
quarks is neutralised.

The analysis of the right-hand diagram of figure~\ref{fig:badjets-app}
is largely similar as long as $k_5$ is at large angles and of the same
hardness as $k_3$ and $k_4$. The additional issue is that now $k_5$
has a collinear divergence with respect to $k_2$. One might generally
worry that semi-hard radiation collinear to $k_2$ might pull $k_3$ far
away from its original direction. This could happen if $k_5$ is
collinear to $k_2$ and if $k_3$ and $k_5$ recombine, with $E_5 \gg E_3$
such that the recombination product ends up collinear to
$k_2$. However if $E_5\gg E_3$ then $y_{35} \gg y_{34}$ and the
$k_3$--$k_4$ will recombine first, neutralising the flavour. Note that if
$E_5 \sim E_3$ and $k_5$ is collinear to $k_2$ then the $k_2-k_5$
recombination will occur first, leaving  the usual (safe) configuration
consisting of a soft $q\bar q$ pair.

One can straightforwardly extend this analysis to multiple $q\bar q$
pairs and multiple gluons. The originally soft large-angle quark can
be dragged further and further towards the hard jet in ensembles with
multiple gluons of similar $k_t$'s but successively larger (but not
strongly ordered) energies.
However, given any fixed number of recombinations, in the soft limit
the resulting quark-like object always has energy $\ll Q$ and will
recombine with the soft antiquark rather than with the hard particles.

A final comment concerns hadron-hadron collisions.  There, the beam
jets have a hardness $k_{tB}(\eta)$, which is of the same order of
magnitude as any hard final-state jets that might have been emitted at
the rapidity $\eta$.  Therefore there is no difference from the point
of view of IRC safety between recombination into final-state jets and
into beam jets and all the arguments given here apply equally well in
the hadron-hadron context.

%======================================================================


\begin{thebibliography}{99}
\bibitem{Spires} It would be excessive to cite them all here. Instead
  the reader may examine them by submitting the spires search
%  "\texttt{T jet AND (T quark OR T gluon) AND PS P}" and then
  ``\texttt{FIND T jet\# AND (T quark OR T gluon)}'' and then
  extracting from the results only those articles whose title contains
  ``quark jet(s)'' or ``gluon jet(s)''.
\end{thebibliography}
\end{document}